\documentclass[twocolumn,superscriptaddress]{revtex4}
\usepackage{graphicx}
\usepackage{subfig}
\usepackage{float}
\usepackage{bm}
\usepackage{color}
\usepackage{amsmath}
\usepackage{amsfonts}
\usepackage[normalem]{ulem}
\usepackage{amssymb}
\usepackage{natbib}
\usepackage{latexsym}
\usepackage{mathrsfs}

\def \ee{\end{equation}}
\def \be{\begin{equation}}
\def \eea{\end{eqnarray}}
\def \bea{\begin{eqnarray}}

\begin{document}

\title{Dark energy as a battery  for magnetic field generation in astrophysical plasmas}

\author{Nicolás Villarroel-Sepúlveda}
\affiliation{Departamento de Física, Facultad de Ciencias, Universidad de Chile, 7800003 Ñuñoa, Santiago, Chile.}

\author{Pablo S. Moya}
\affiliation{Departamento de Física, Facultad de Ciencias, Universidad de Chile, 7800003 Ñuñoa, Santiago, Chile.}

\author{Felipe A. Asenjo} \email{felipe.asenjo@uai.cl}%
\affiliation{Facultad de Ingeniería y Ciencias, Universidad Adolfo Ibáñez,  Peñalol\'en 7941169, Santiago, Chile. }%

\author{Swadesh M. Mahajan}
\affiliation{Institute for Fusion Studies, The University of Texas at Austin, Texas 78712, USA.}

\date{\today}

\begin{abstract}
We show that in the spacetime dominated by a cosmological constant, in the far region of a  Schwarzschild-de Sitter black hole, a seed magnetic field can be generated in an ambient plasma (in a state of no magnetic field) by a general-relativistic battery. This cosmological battery depends on the interaction of spacetime curvature with inhomogeneous plasma thermodynamics. Thus, dark energy becomes the only gravitational source for magnetic field generation at large distances from the black hole. This allows a mechanism that makes dark energy manifest through its conversion to cosmic magnetic fields.  
\end{abstract}

\maketitle


When the primary source of magnetic fields - the electrical current - is not explicitly evident in a system, one must resort to indirect sources with free energy that could generate magnetic energy. There is a large body of literature on the so-called dynamo mechanisms (see, for instance, Refs.~\cite{KR, MGM, MGM2, MS}) that could amplify a magnetic field from a non-zero seed. Therefore, looking for processes/systems that could generate a seed magnetic field has been a serious quest in plasma physics. Let us call such seed creation systems, collectively, Batteries. The most famous of these is the very original Biermann Battery (BB) \cite{biermann,kulr}. The seed magnetic field in the BB originates in the inhomogeneous thermodynamics (of a special type) that harnesses the free energy in density and temperature gradients. 

In principle, the Biermann battery can exist for plasmas at all energy scales. However, it is neither unique nor always dominant; when special and/or general relativistic effects are considered, new batteries appear \cite{yoshida,asenjoMahajan,M2,asenjoMahajan2} and also may dominate. These batteries may originate from different free energy sources. For instance, the source of magnetic energy in Refs.~\cite{yoshida,M2} lies in the free energy in velocity shears. In addition, in Refs.~\cite{asenjoMahajan,asenjoMahajan2}, the possibility of using the spacetime curvature as the source for magnetic field generation is explored. These relativistic batteries are likely more effective in ``extreme conditions".

In this work, we explore precisely one of these systems, the plasma dynamics in the background of a Schwarzschild-de Sitter black hole. We plan to demonstrate the working of the general relativistic battery \cite{asenjoMahajan} driven by the interaction of this spacetime curvature with plasma thermodynamics. In particular, we will focus on the vortical plasma dynamics very far from the black hole - in the region where the spacetime curvature is controlled entirely by the cosmological constant (dark energy), as the curvature produced by the black hole mass is negligible. To our knowledge, this is the first calculation connecting dark energy to magnetic field generation (seed).

The spacetime  curvature of the Schwarzschild-de Sitter black hole \cite{rindler} is described by the metric $g_{\mu\nu}$; its non-vanishing components (in spherical coordinates) are 
$g_{tt}=-\alpha^{2}$, $g_{rr}=\alpha^{-2}$, $g_{\theta\theta}=r^2$, and $g_{\phi\phi}=r^2\sin^2\theta$,
with 
\begin{equation}\label{metricScdS}
    \alpha=\left(1-\frac{2 M}{r}-\frac{\Lambda r^2}{3}\right)^{1/2}\, ,
\end{equation} 
where $M$ is the black hole mass, and $\Lambda$ is the positive cosmological constant of a  de Sitter Universe. This metric allows the realization of a black hole in an expanding Universe. In this spacetime, several plasma propagation modes have been studied \cite{Hossain,sharif,Atiqur}.

In this work, we study the dynamics of a plasma on the spacetime background given by the previous metric. In general, the covariant dynamics of a non-gravitating plasma fluid (the intrinsic gravity of the plasma is neglected) in a curved spacetime background can be described in terms of a unified field $M^{\mu\nu}$ \cite{M1-M3,asenjoMahajan}. The dynamics can be put in the unified form
\begin{equation}
U_\nu M^{\mu\nu}=\frac{T}{q}\nabla^\mu \sigma\, ,
\label{ecprinci}
\end{equation}
where $M^{\mu\nu}=F^{\mu\nu}+(m/q) S^{\mu\nu}$ 
is a fully antisymmetric tensor comprising the electromagnetic tensor $F^{\mu\nu}$, and $\nabla_\mu$ is the covariant derivative defined for a general metric $g_{\mu\nu}$. The plasma, composed by particles with mass $m$, and charge $q$, is characterized by the tensor $S^{\mu\nu}=\nabla^\mu (f U^\nu)-\nabla^\nu (f U^\mu)$, with four-velocity $U^\mu$, and temperature $T$. Besides, $\sigma$ and $f$ are the entropy and enthalpy per mass density, respectively \cite{M1-M3}. 
For a relativistic Maxwell distribution, $f(x)=K_3(x)/K_2(x)$, where $K_j$
is the modified Bessel function of order $j$, and $x=mc^2/k_BT$ is the inverse normalized temperature, with the Boltzmann constant $k_B$ \cite{M1-M3}.
This entropy is related to pressure $p$ and enthalpy by $\nabla^\mu\sigma=\nabla^\mu p/nT-m\nabla^\mu f/T$.
The dynamics is complete when complemented by the generally covariant Maxwell equations. 

From the basic Eq.~\eqref{ecprinci}, the expression for magnetic field generation (in a spherically symmetric static spacetime) can be readily derived by following the procedure detailed in Ref.~\cite{asenjoMahajan}. One affects a $3+1$ decomposition of spacetime, projecting all physical quantities onto timelike and spacelike hypersurfaces; the spacetime metric is decomposed as $g_{\mu\nu}=\gamma_{\mu\nu}-n_\mu n_\nu$, where $n_\mu$ is a normalized timelike vector that satisfies $n_\mu n^\mu=-1$, and  $n^\mu\gamma_{\mu\nu}=0$. Similarly, $\gamma_{\mu\nu}$ is the 3-metric of the spacelike hypersurfaces of metric $g_{\mu\nu}$. For the spherically symmetric static spacetime of the Schwarzschild-de Sitter black hole, the metric acquires the form $g_{\mu\nu}=(-\alpha^2, \gamma_{ij})$, yielding 
$n_\mu=(\alpha,0,0,0)$, $\gamma_{rr}=\alpha^{-2}$, $\gamma_{\theta\theta}=r^2$, and $\gamma_{\phi\phi}=r^2\sin^2\theta.$
By combining  the spacelike projection of Eq.~\eqref{ecprinci} with the spacelike projection of the identity $\nabla_\nu M^{*\mu\nu}=0$ (where $M^{*\mu\nu}$ is the dual of $M^{\mu\nu}$),  the generalized plasma vorticity equation in curved spacetimes is found to be \cite{asenjoMahajan}
\begin{equation}
    \frac{\partial {\bf \Omega}}{\partial t}-\nabla\times\left({\bf v}\times{\bf\Omega}\right)={\bf \Xi}_B+{\bf \Xi}_R\, ,
\label{generalizedVEc}
\end{equation}
where the  generalized plasma vorticity ${\bf \Omega}$ is  the vectorial part of the spacelike tensor $\Omega^\mu=(1/2)n_\rho \epsilon^{\rho \mu\sigma\tau}M_{\sigma\tau}$, with the totally antisymmetric tensor $\epsilon^{\rho \mu\sigma\tau}$ (notice that $n_\mu \Omega^\mu=0$).
Also,  
${\bf v}$ is the plasma fluid velocity, being the vectorial part of the four-velocity  $v^\alpha=(1/\Gamma){\gamma^\alpha}_\mu U^\mu$. Here, $\Gamma=(1/\alpha)n_\mu U^\mu$ is the corresponding Lorentz factor of the fluid velocity, such that the four-velocity can be written as $U^\mu=-\alpha \Gamma n^\mu+\Gamma {\gamma^\mu}_\nu v^\nu$. Thus, $\Gamma=(\alpha^2-\gamma_{\mu\nu}v^\mu v^\nu)^{-1/2}$. Besides, the $\nabla$ operator corresponds to the spacelike projection of the covariant derivative. Furthermore, the vectorial components of the generalized plasma vorticity
can be written as
\begin{equation}
    {\bf \Omega}={\bf B}+\frac{m}{q}\nabla\times\left(f\Gamma{\bf v}\right)\, ,
    \label{generalizedV}
\end{equation}
where ${\bf B}$ is the vectorial part of the spacelike  magnetic field tensor, defined as 
\begin{equation}\label{magfielddef}
B^\mu=\frac{1}{2}n_\rho \epsilon^{\rho \mu\sigma\tau}F_{\sigma\tau}\, .
\end{equation}
Also, the term $\nabla\times\left(f\Gamma{\bf v}\right)$ in Eq.~\eqref{generalizedV} corresponds to the vectorial part of the fluid vorticity tensor.

The two terms on the right-hand side of Eq.~\eqref{generalizedVEc} are the batteries of the theory. This equation predicts that these batteries  generate generalized vorticity.
The first one is the general relativistic-corrected Biermann battery \cite{asenjoMahajan}
\begin{equation}
    {\bf \Xi}_B=-\frac{1}{q\Gamma}\nabla T\times\nabla \sigma\, ,
    \label{bateryB}
\end{equation}
that depends only on the non-parallel spatial variations of the plasma temperature and entropy. In most simple cases, both quantities tend to have parallel gradients, and the Biermann battery vanishes. The second battery  
\begin{equation}
    {\bf \Xi}_R=\frac{T}{q\Gamma^2}\nabla \Gamma\times\nabla \sigma\, ,
    \label{bateryR},
\end{equation}
is the general relativistic drive \cite{asenjoMahajan}, the prime objective of this paper. Notice that this battery appears by the interplay between relativistic kinematical effects, spacetime curvature, and the thermodynamical properties of the plasma; the spacetime curvature is  embedded in the definition of $\Gamma$.

In this unified formalism, the electromagnetic field and the fluid thermal-vortical field appear together, and what is being generated via Eq.~\eqref{generalizedVEc} is the generalized vorticity (or the generalized magnetic field). In the conventional non-relativistic theories like the original BB, the thermal-vortical part is neglected, and the generalized vorticity is reduced to the magnetic field \cite{M1-M3}.

The generalized vorticity equation \eqref{generalizedVEc} and the importance of the general relativistic drive shown in Eq.~\eqref{bateryR} have been explored in several astrophysical scenarios, such as for plasmas around Schwarzschild black holes \cite{asenjoMahajan,kana},
Reissner-N\"ordstrom and Kerr black holes \cite{asenjoMahajan2,Bhattacharjee1,Bhattacharjee2,Bhattacharjee3,Bhattacharjee4}. 
In all those cases, the mass of the black hole is the source of the space-time curvature; the magnetic field generation is, thereby, a local phenomenon. Here, we make a fundamental departure in exploring the effect of the pervading dark energy through the cosmological constant.

To calculate the effect of the cosmological constant on the generation of the magnetic field seed, and in the same spirit as the earlier general relativistic calculations, we will study the conditions in which the general relativistic drive \eqref{bateryR} is more relevant than the Biermann battery \eqref{bateryB}. Without loss of generality, we consider the metric \eqref{metricScdS}, evaluated at the black hole's equatorial plane ($\theta=\pi/2$), to study the plasma dynamics.
Far from the black hole, let us assume that the fluid has only radial velocity, ${\bf v}=v^r \hat e_r$, and $v^r=v^r(r)$. For this choice, the fluid vorticity vanishes identically, and the generalized vorticity \eqref{generalizedV} reduces to the magnetic field, ${\bf \Omega}={\bf B}$.

Due to the chosen symmetry of our system, let us consider the generation of a magnetic field with a polar component in the equatorial plane. The components of the magnetic field can be explicitly evaluated using Eq.~\eqref{magfielddef}. This allows us to obtain that the orthogonal component to the equatorial plane is ${B}^\theta=\alpha B^\theta_{\mbox{\footnotesize{flat}}}$, where
$B^\theta_{\mbox{\footnotesize{flat}}}$ is the polar magnetic field (in the equatorial plane) in flat spherical coordinates. We have explicitly isolated the effect of curvature in the magnetic field.
Thereby, the (only non-vanishing) polar component of Eq.~\eqref{generalizedVEc} can be written as
\begin{equation}
    \frac{\partial {B}^\theta_{\mbox{\footnotesize{flat}}}}{\partial t}+\frac{1}{r^2}\frac{\partial}{\partial r}\left({r^2}v^r B^\theta_{\mbox{\footnotesize{flat}}}\right)=\frac{{ \Xi}_B^\theta}{\alpha}+\frac{{\Xi}_R^\theta}{\alpha}\, .
\label{generalizedVEcb}
\end{equation}

In order to evaluate the contribution of the general relativistic drive, let us first model the simplest case where the Biermann battery \eqref{bateryB} vanishes. This occurs when $\nabla T\parallel \nabla \sigma$. In such cases, the only available battery will be the one given by \eqref{bateryR}. To evaluate this battery, let us consider an initial state with no plasma and no magnetic field. Thus, we have an initial flow of particle constituents moving radially, with velocity  
$v^r=U^r/\Gamma=\dot r/\Gamma=\sqrt{E^2-\alpha^2}/\Gamma$ \cite{jak}, where $E$ is a dimensionless constant related to the energy density of the \textcolor{red}{fluid element} \sout{particle}, determined at the point $r_0$ when the particle is at rest, $E=\alpha(r_0)$]. Besides, the Lorentz factor is given by $\Gamma={E}/{\alpha^2}$ \cite{jak}. With this, we can calculate the relativistic drive \eqref{bateryR}, which is non-zero only if angular gradients of the entropy are allowed. The polar component of the relativistic drive, then, is
\begin{equation}
    {\Xi}_R^\theta=\frac{T\alpha^5}{q E r}\frac{\partial}{\partial r}\left(\frac{1}{\alpha^2}\right)\frac{\partial \sigma}{\partial\phi}\, ,
\end{equation}
which, very far from the black hole, becomes
\begin{equation}
    \left.\frac{{\Xi}_R^\theta}{\alpha}\right|_{r\ggg 2M}\approx\frac{2 T\, \Lambda}{3q E}\frac{\partial \sigma}{\partial\phi}\, ,
    \label{relatdrieinfinity}
\end{equation}
scaling linearly with the cosmological constant  $\Lambda$. Notice that, for this battery not to diverge at large distances, the temperature $T$ and entropy $\sigma$ must remain bounded. Also, this battery does not exist in the flat spacetime limit.

In contrast, one may wonder under what conditions the Biermann battery can compete with the cosmological constant induced relativistic drive shown in Eq.~\eqref{relatdrieinfinity}. If the temperature and entropy gradients are not parallel, the polar component of the Biermann battery \eqref{bateryB} is ${\Xi}_B^\theta=-({\alpha^3}/{q E r}){\partial_r T} {\partial_\phi \sigma}$, which, far away from the black hole, becomes 
\begin{equation}
  \left.\frac{|{\bf \Xi}_B^\theta|}{\alpha}\right|_{r\ggg 2M}\approx \frac{\left(-1+\Lambda r^2/3\right)}{3q Er}\frac{\partial T}{\partial r} \frac{\partial \sigma}{\partial \phi}\, .
  \label{biermannfar}
 \end{equation}

The result \eqref{biermannfar} has the correct flat spacetime limit of the Biermann battery in the case of vanishing cosmological constant. However, when the cosmological 
constant $\Lambda$ is non-vanihing and it has a very small value, the lapse function \eqref{metricScdS} is defined only to large distances of the order
\begin{equation}
    r\approx\sqrt{\frac{3}{\Lambda}}\, .
    \label{veryfardustacer}
\end{equation}
 This implies that, for a plasma very far from a  Schwarzschild-de Sitter black hole, at distances of the order of \eqref{veryfardustacer}, the Biermann battery \eqref{biermannfar} is completely negligible. Thus, the dark energy-dominated part of the general relativistic drive will dominate the generation of magnetic field. Let us name this drive as the Dark Energy Battery (DEB).

Let us go back to Eq.~\eqref{generalizedVEcb} to estimate the magnetic field generated by DEB starting from a state of no magnetic field. For a general time evolution, the calculation will involve solving complicated nonlinear equations. In fact, a self-consistent solution can be affected by solving Eq.~\eqref{generalizedVEcb}, considering the dynamics of plasma entropy. Such work will necessarily involve numerical solutions of the system and will be deferred to a later publication. This paper, devoted to the first recognition of a new pervading drive, will do the local estimate of the linear phase of the seed generation. 
 In this regime, because battery \eqref{relatdrieinfinity}, only the cosmological constant is responsible for generating magnetic fields. In the very-far limit ($r\ggg 2M)$, for times measured from an arbitrary $t=0$, the rate of magnetic field generation is
 \begin{equation}
    \frac{\partial}{\partial t}{ B^\theta_{\mbox{\footnotesize{flat}}}(r\ggg 2M)}\approx \frac{2T\Lambda}{3qE} \frac{\partial \sigma}{\partial \phi}\, ,
    \label{campomaggenb}
\end{equation}
which remains well-defined as the thermodynamic quantities remain finite. Notice that the only gravitational (source of spacetime curvature) entity in Eq.~\eqref{campomaggenb} is the cosmological constant (dark energy).  

Eq.~\eqref{campomaggenb} gives the non-vanishing initial condition required to generate a magnetic field. Thus, the magnetic field grows linearly with time until a time $\tau$, when the nonlinearities of the vortical dynamics [second term at the left-hand side of Eq.~\eqref{generalizedVEcb}] become relevant. This time can be estimated to be $\tau\sim |L/v^r|$, where $L$ is the length scale of radial variations of the magnetic field \cite{asenjoMahajan}. The condition for the radial variations of $B^\theta_{\mbox{\footnotesize{flat}}}$ to be relevant in this spacetime is $1/L\sim |\partial_r B^\theta_{\mbox{\footnotesize{flat}}}/B^\theta_{\mbox{\footnotesize{flat}}}|\sim |\partial_r\alpha/\alpha|$. For the metric \eqref{metricScdS}, very far from the black hole, we find that $L\sim r$, implying that magnetic fields present almost no variations at infinity. With this, at large distances, the time scale (far from the black hole), then, goes as
\begin{equation}
   \left.{\tau}\right|_{r\ggg 2M}\sim \frac{E}{r^2}\left(\frac{3}{\Lambda} \right)^{3/2}\, .
\end{equation}

At time $\left.{\tau}\right|_{r\ggg 2M}$, the magnetic field will grow to its maximum value (permitted in this model) given by  
\begin{equation}
B^\theta_{\mbox{\footnotesize{max}}}(r\ggg 2M)\approx \frac{2T}{q\,  r^2}\sqrt{\frac{3}{\Lambda}} \frac{\partial \sigma}{\partial \phi}\approx \frac{2T}{q\,  }\sqrt{\frac{\Lambda}{3}} \frac{\partial \sigma}{\partial \phi}\, ,
     \label{maxmagfield}
\end{equation}
where in the last approximation we have used the maximum distance estimate \eqref{veryfardustacer}. Thus, very far from the black hole, the generated polar (angular) magnetic field depends only on the plasma thermodynamics and dark energy.


The above simplest model calculation shows that the cosmological constant (pervading dark energy) can trigger the generation of a magnetic field \eqref{maxmagfield} when a plasma is present in the outer reaches of a black hole, with local variations on its entropy density. Like the conventional BB battery, this new battery is powered by the interaction of gradient-free energy, this time with the space-time curvature induced by the dark energy. The linear growth stage (from an initial state of no magnetic field) lasts for a time $\tau$ during which the field reaches a maximum value \eqref{maxmagfield}. About and beyond the time $\tau$, a much more elaborate calculation is needed.


The $B^\theta_{\mbox{\footnotesize{max}}}$ generated by this mechanism (in the region very far from the black hole) is expected to be rather small. The value of the cosmological constant $\Lambda$ is very small, as well as the azimuthal variations of entropy (and, therefore, of temperature). By considering a cold plasma ($f\approx 1$), then  $\partial_\phi\sigma\approx\partial_\phi p/nT\equiv\chi\ll 1$ is a small dimensionless quantity. Therefore, from magnetic field \eqref{maxmagfield}, we can estimate its strength compared to the square root of the rest mass energy density of the plasma constituents as
\begin{equation}
\frac{|B^\theta_{\mbox{\footnotesize{max}}}|}{\sqrt{nm}}=\left( \frac{\lambda_D^2\sqrt{\Lambda}}{\lambda_p}\right)\chi\, ,
\label{finalestimation}
\end{equation}
where $\lambda_D=v_{th}/\omega_p$ is the Debye length (with the plasma thermal velocity $v_{th}$ and the plasma frequency $\omega_p$), and $\lambda_p=1/\omega_p$ is the plasma inertial length. Thus, the generated magnetic field depends on the interaction between the characteristic lengths of the plasma ($\lambda_D$ and $\lambda_p$) and the main length of a spacetime with dark energy (${\Lambda}^{-1/2}$). By considering  an interstellar electron plasma, with density $n\sim 0.1$ cm$^{-3}$ and temperature $T\sim 10$ eV \cite{chen}, then $\lambda_D\sim 74$ m, and $\lambda_p\sim 1.7\times 10^4$ m. Therefore, using a conservative estimation for $\Lambda\sim 10^{-52}$ m$^{-2}$ \cite{solnic}, we obtain from Eq.~\eqref{finalestimation} that $|B^\theta_{\mbox{\footnotesize{max}}}|/{\sqrt{nm}}\sim 3\chi\times 10^{-27}\lll 1$. Therefore, the magnetic field seed has an energy that is several tens of orders of magnitude less than the rest mass energy of the plasma constituents. However, this seed can later be amplified by other varied mechanisms.

Finally, it is to be emphasized that, in the spacetime considered here (dominated by dark energy), the general relativistic drive DEB \eqref{relatdrieinfinity} completely dominates the Biermann battery and is the primary cause of a finite magnetic field seed \eqref{campomaggenb}. This is the principal result of this work, where we explored a unique new mechanism where dark energy can be converted into magnetic energy. This calculation establishes a basis (in principle) for detecting the signatures of dark energy through electromagnetic energy. By the same token, if there is dark energy, we will always have a source for cosmic magnetic fields.  

The very idealized simple estimates made in this paper need to be supplemented by a solution of the nonlinear Eq.~\eqref{generalizedVEc} [or Eq.~\eqref{generalizedVEcb}] to get a better idea about the saturation of the magnetic field. This will be allowed by fully coupling the dynamics to the Maxwell equations. Such a system will have to be solved numerically and will be the subject of a more detailed paper. 
The above simple analysis, however, has provided ample support for the mechanism that may convert the dark energy into well-known and well-understood magnetic energy.

\section*{Acknowledgments}
We are grateful to ANID-Chile for   Doctoral Scholarship No. 21220616 (NVS), and Fondecyt grants No. 1240281 (PSM) and No. 1230094 (FAA), which partially supported this work. SMM's research is supported by US DOE Grants DE-FG02-04ER54742 and DE-AC02-09CH11466.

\end{document}